\documentclass[sn-mathphys,Numbered]{sn-jnl}

\usepackage{graphicx}%
\usepackage{multirow}%
\usepackage{amsmath,amssymb,amsfonts}%
\usepackage{amsthm}%
\usepackage{mathrsfs}%
\usepackage[title]{appendix}%
\usepackage{xcolor}%
\usepackage{textcomp}%
\usepackage{manyfoot}%
\usepackage{booktabs}%
\usepackage{algorithm}%
\usepackage{algorithmicx}%
\usepackage{algpseudocode}%
\usepackage{listings}%
\newcommand{\eplus}{\ensuremath{\mathrm{e^+}}}
\newcommand{\pbar}{\ensuremath{\mathrm{\overline{p}}}}
\newcommand{\Hbar}{\ensuremath{\mathrm{\overline{H}}}}
\newcommand{\Hbarplus}{\ensuremath{\mathrm{\overline{H}^+}}}
\newcommand{\Hminus}{\ensuremath{\mathrm{H}^-}}

\begin{document}

\title[Article Title]{Production of antihydrogen atoms by 6~keV antiprotons through a positronium cloud}

\author[1]{\fnm{P.} \sur{Adrich}} 
\author[2]{\fnm{P.} \sur{Blumer}}
\author[2]{\fnm{G.} \sur{Caratsch}}
\author[3]{\fnm{M.} \sur{Chung}}
\author[4]{\fnm{P.} \sur{Cladé}}
\author[5]{\fnm{P.} \sur{Comini}}
\author[2]{\fnm{P.} \sur{Crivelli}}
\author[6]{\fnm{O.} \sur{Dalkarov}}
\author[5]{\fnm{P.} \sur{Debu}}
\author[4,7]{\fnm{A.} \sur{Douillet}}
\author[4]{\fnm{D.} \sur{Drapier}}
\author[8,20]{\fnm{P.} \sur{Froelich}}
\author[4,21]{\fnm{N.} \sur{Garroum}}
\author[4,9]{\fnm{S.} \sur{Guellati-Khelifa}}
\author[4]{\fnm{J.} \sur{Guyomard}}
\author[10]{\fnm{P-A.} \sur{Hervieux}}
\author[4,7]{\fnm{L.} \sur{Hilico}}
\author[4]{\fnm{P.} \sur{Indelicato}}
\author[8]{\fnm{S.} \sur{Jonsell}}
\author[4,7]{\fnm{J-P.} \sur{Karr}}
\author[11]{\fnm{B.} \sur{Kim}}
\author[12]{\fnm{S.} \sur{Kim}}
\author[13]{\fnm{E-S.} \sur{Kim}}
\author[11]{\fnm{Y.J.} \sur{Ko}}
\author[1]{\fnm{T.} \sur{Kosinski}}
\author[14]{\fnm{N.} \sur{Kuroda}}
\author[5,22]{\fnm{B.M.}\sur{Latacz}}
\author[12]{\fnm{B.} \sur{Lee}}
\author[12]{\fnm{H.} \sur{Lee}}
\author[11]{\fnm{J.} \sur{Lee}}
\author[13]{\fnm{E.} \sur{Lim}}
\author[5]{\fnm{L.} \sur{Liszkay}}
\author[15]{\fnm{D.} \sur{Lunney}}
\author[10]{\fnm{G.} \sur{Manfredi}}
\author[5]{\fnm{B.} \sur{Mansoulié}}
\author[1]{\fnm{M.} \sur{Matusiak}}
\author[16]{\fnm{V.} \sur{Nesvizhevsky}}
\author[4]{\fnm{F.} \sur{Nez}}
\author[15,22]{\fnm{S.} \sur{Niang}}
\author[2]{\fnm{B.} \sur{Ohayon}}
\author[11,12]{\fnm{K.} \sur{Park}}
\author[4]{\fnm{N.} \sur{Paul}}
\author[5]{\fnm{P.} \sur{Pérez}}
\author[2]{\fnm{C.} \sur{Regenfus}}
\author[4]{\fnm{S.} \sur{Reynaud}}
\author[15]{\fnm{C.} \sur{Roumegou}}
\author[5]{\fnm{J-Y.} \sur{Roussé}}
\author[5]{\fnm{Y.} \sur{Sacquin}}
\author[5]{\fnm{G.} \sur{Sadowski}}
\author[2]{\fnm{J.} \sur{Sarkisyan}}
\author[14]{\fnm{M.} \sur{Sato}}
\author[17]{\fnm{F.} \sur{Schmidt-Kaler}}
\author[1]{\fnm{M.} \sur{Staszczak}}
\author[1]{\fnm{K.} \sur{Szymczyk}}
\author[14]{\fnm{T.A.} \sur{Tanaka}}
\author[5]{\fnm{B.} \sur{Tuchming}}
\author[5]{\fnm{B.} \sur{Vallage}}
\author[6]{\fnm{A.} \sur{Voronin}}
\author[18]{\fnm{D.P.} \sur{van der Werf}}
\author[12]{\fnm{D.} \sur{Won}}
\author[1]{\fnm{S.} \sur{Wronka}}
\author[19]{\fnm{Y.} \sur{Yamazaki}}
\author[3]{\fnm{K-H.} \sur{Yoo}}
\author[4]{\fnm{P.} \sur{Yzombard}}

\affil[1]{\orgdiv{National Centre for Nuclear Research}, \orgname{(NCBJ)}, \orgaddress{\street{ul. Andrzeja Soltana 7}, \postcode{} \city{05-400 Otwock, Swierk},  \country{Poland}}}
\affil[2]{\orgdiv{Institute for Particle Physics and Astrophysics}, \orgname{ETH Zurich}, \orgaddress{ \city{8093 Zurich}, \country{Switzerland}}}
\affil[3]{\orgdiv{Department of Physics}, \orgname{Ulsan National Institute of Science and Technology (UNIST)}, \orgaddress{\street{50, UNIST-gil}, \city{Ulsan}, \postcode{44919}, \country{Republic of Korea}}}
\affil[4]{\orgdiv{Laboratoire Kastler Brossel}, \orgname{Sorbonne Université, CNRS, ENS-Université PSL, Collège de France, Campus Pierre et Marie Curie}, \orgaddress{\street{4, place Jussieu}, \postcode{F-75005},\city{Paris},  \country{France}}}
\affil[5]{\orgdiv{IRFU}, \orgname{CEA, Université Paris-Saclay}, \orgaddress{\street{} \postcode{F-91191}, \city{Gif-sur-Yvette},  \country{France}}}
\affil[6]{Affiliated with an institute covered by a cooperation agreement with CERN}
\affil[7]{\orgdiv{} \orgname{Université d'Evry-Val d'Essonne, Université Paris-Saclay}, \orgaddress{\street{Boulevard François Mitterand}, \postcode{F-91000}, \city{Evry},  \country{France}}}
\affil[8]{\orgdiv{Department of Physics}, \orgname{ Stockholm University}, \orgaddress{\postcode{SE-10691}, \city{Stockholm},  \country{Sweden}}}
\affil[9]{\orgname{Conservatoire National des Arts et Métiers}, \orgaddress{\street{292 rue Saint Martin}, \postcode{F-75003}, \city{Paris}, \country{France}}}
\affil[10]{\orgname{Université de Strasbourg, CNRS}, \orgdiv{Institut de Physique et Chimie des Matériaux de Strasbourg UMR 7504}, \orgaddress{\postcode{F-67000}, \city{Strasbourg}, \country{France}}}
\affil[11]{\orgname{Center for Underground Physics}, \orgdiv{Institute for  Basic Science}, \orgaddress{\street{70 Yuseong-daero 1689-gil,  Yuseong-gu}, \city{Daejeon 34047}, \country{Republic of Korea}}}
\affil[12]{\orgdiv{Department of Physics and Astronomy}, \orgname{Seoul National University}, \orgaddress{\street{1 Gwanak-Ro, Gwanak-gu}, \city{Seoul 08826}, \country{Republic of Korea}}}
\affil[13]{\orgdiv{Department of Accelerator Science}, \orgname{Korea University Sejong Campus}, \orgaddress{\street{Sejong-ro 2511}, \postcode{30019} \city{Sejong}, \country{Republic of Korea}}}
\affil[14]{\orgdiv{Institute of Physics}, \orgname{University of Tokyo}, \orgaddress{\street{3-8-1 Komaba, Meguro}, \city{Tokyo} \postcode{153-8902}, \country{Japan}}}
\affil[15]{\orgname{Université Paris-Saclay, CNRS/IN2P3}, \orgdiv{IJCLab}, \orgaddress{\city{Orsay}, \country{France}}}
\affil[16]{\orgname{Institut Max von Laue - Paul Langevin (ILL)}, \orgaddress{\street{71 avenue des Martyrs}, \city{Grenoble}, \country{France}, \postcode{F-38042}}}
\affil[17]{\orgdiv{QUANTUM, Institut für Physik}, \orgname{Johannes Gutenberg Universität}, \orgaddress{\postcode{D-55128} \city{Mainz}, \country{Germany}}}
\affil[18]{\orgdiv{Department of Physics, Faculty of Science and Engineering}, \orgname{Swansea University}, \orgaddress{\city{Swansea} \postcode{SA2 8PP}, \country{United Kingdom}}}
\affil[19]{\orgdiv{Ulmer Fundamental Symmetries Laboratory}, \orgname{RIKEN}, \orgaddress{\street{2-1 Hirosawa}, \city{Wako}, \postcode{351-0198}, \state{Saitama}, \country{Japan}}}
\affil[20]{\orgdiv{Present address}, \orgname{Uppsala University, \AA ngstr\"om Laboratory, Department of Chemistry}, \orgaddress{\street{Box 523}, \postcode{75120} \city{Uppsala}, \country{Sweden}}}
\affil[21]{\orgdiv{Present address}, \orgname{LPNHE/IN2P3}, \orgaddress{\street{4, place Jussieu}, \postcode{75252} \city{Paris}, \country{France}}}
\affil[22]{\orgdiv{Present address}, \orgname{CERN}, \orgaddress{\street{Esplanade des Particules 1}, \postcode{1217} \city{Meyrin}, \country{Switzerland}}}

\abstract{
We report on the first production of an antihydrogen beam by charge exchange of 6.1~keV antiprotons with a cloud of positronium in the GBAR experiment at CERN. The antiproton beam was delivered by the AD/ELENA facility. The positronium target was produced from a positron beam itself obtained from an electron linear accelerator. We observe an excess over background indicating antihydrogen production with a significance of 3-4 standard deviations.  
}

\maketitle

\section{Introduction}
\label{sec1}

The GBAR experiment at CERN aims at a precise measurement of the free fall acceleration of neutral antihydrogen atoms in the terrestrial gravitational field, thus testing the Weak Equivalence Principle with antimatter. 
The antimatter atoms must be cooled to $\mu $K temperatures (i.e. $\mathrm{\sim 1 \; m s^{-1}}$ velocities) to measure their free fall. 
The plan is to follow the original idea of J.~Walz and T.~Hänsch~\cite{Walz}. In this scheme, an anti-ion (~\Hbarplus) is first produced  and sympathetically cooled with laser cooled $\mathrm{Be^+}$ ions.  After photo-detachment at threshold, an ultra-cold anti-atom is produced. In order to obtain the anti-ions, two consecutive charge exchange reactions on positronium (Ps) are needed. In the first reaction, an antiproton beam interacts with a cloud of Ps to produce antihydrogen atoms, whereas in the second reaction those anti-atoms interact with another Ps from the cloud to form the antihydrogen ion~\cite{proposal}.

In this work we present the experimental realisation of
the  first step of this scheme where a beam of antiprotons of 6.1 keV energy interacts with a cloud of Ps in the fundamental state, producing anti-atoms according to reaction:
\begin{equation}
\mathrm{\overline{p}+Ps \rightarrow \overline{H} + e^-} 
\label{eq1}
\end{equation}
 where \pbar\ stands for antiproton and \Hbar\ for antihydrogen.
 
The production of antihydrogen was first demonstrated in 1996 by the PS210 experiment at the CERN LEAR ring where 1.94 GeV/c antiprotons produced $e^+e^-$ pairs in a Xe  target: 
$\mathrm{\overline{p}  + Z \rightarrow \overline{p} \; \gamma \gamma \;Z \rightarrow \overline{p} \; e^+e^- \; Z \rightarrow \overline{H} \; e^- \; Z}$~\cite{PS210}. 
In 1998, the E862 experiment  at Fermilab used a high energy beam of  5.2 - 6.2 GeV/c antiprotons on a hydrogen gas jet target and the reaction: $\mathrm{p + \overline{p} \rightarrow \overline{H} + e^- + p}$~\cite{E862}. 
These experiments  produced a few high energy anti-atoms.
In 2002, the ATHENA and ATRAP collaborations at the CERN antiproton decelerator facility succeeded 
in producing sub-eV antihydrogen atoms~\cite{ATHENA,ATRAP0}.
In 2010 the ALPHA collaboration was able to catch about 38 \Hbar s in a magnetic octupole trap, superimposed on a one tesla magnetic field \cite{ALPHATrap1}. The trap depth for ground state antihydrogen was about 0.5~K.
The same year, the ASACUSA collaboration reported production using a CUSP trap~\cite{ASACUSA} in order to produce a very low energy beam of \Hbar\ (50 K temperature), which has been partially demonstrated~\cite{ASACUSA2}.\\
The main formation mechanism involved  in these trap experiments is a three-body reaction with one antiproton and two positrons, the second positron
carrying away the excess energy released from the formation of the antihydrogen atom. Such a process is very efficient at producing several tens  of anti-atoms per cycle that can be trapped as shown by the ALPHA collaboration~\cite{ALPHA}.
In 2004 the ATRAP collaboration demonstrated another method involving positronium (Ps) itself produced by the interaction of positrons with excited Cesium atoms, in two resonant charge exchange reactions: $\mathrm{Cs^* + e^+ \rightarrow Ps^* + Cs^+}$ followed by $\mathrm{Ps^* + \overline{p} \rightarrow \overline{H}^* + e^-}$~\cite{ATRAP}. Here the third body is the electron from Ps. 
In 2021 the AEgIS collaboration reported the pulsed production of antihydrogen atoms in Rydberg states via the interaction of trapped antiprotons with Rydberg Ps~\cite{AEGIS}: 
$\mathrm{Ps^* + \overline{p} \rightarrow \overline{H}^* + e^-}$. \\

In the experiment described here, the Ps atoms are in their fundamental state, thus producing  \Hbar\ in low excitation states. The anti-atoms are not trapped but produced in the form of a pulsed beam.

The  charge conjugated process of reaction~(\ref{eq1}) 
was measured in 1997 using a beam of 11.3 to 15.8 keV protons colliding with Ps~\cite{Merrison} and producing hydrogen atoms with cross sections in the 
$\mathrm{10^{-15}\; cm^2}$ range.
However, the precision on the measurement did not
allow to settle the disagreement between the various atomic-physics models reported in~\cite{Merrison}. 
These models, along with more recent calculations~\cite{Comini,Leveque-Hervieux,Kadyrov}, predict
in any case that the cross sections should be higher at lower energy and exhibit a maximum in the
range 6 to 10~keV. Moreover, the production of antihydrogen ions, which is the next step of the GBAR scheme, is most efficient when the antihydrogen is in its
ground state. The corresponding cross sections are also higher at lower energy, towards the reaction thresholds~\cite{secondreaction}. Assuming that positronium and antihydrogen are in their ground states, the threshold is 5.6~keV and this is a further motivation for reducing the antiproton beam energy close to 6~keV.

Reaction~(\ref{eq1}) with ground state positronium  produces antihydrogen atoms in low excited states, with 14\%~\cite{Leveque-Hervieux} to 22\%~\cite{Kadyrov} directly in the ground state, and the majority
in the 2P state that  rapidly decay to the ground state (lifetime of 1.6~ns). Compared to antiproton mixing with positrons in nested traps or with highly excited states of Ps, the present method is well suited for ground state antihydrogen production and therefore to subsequent antihydrogen ion formation.\\

In the following sections we first describe the main parts of the apparatus, and how we determine the number of Ps  atoms and antiprotons participating in reaction~(\ref{eq1}) for each ELENA pulse. We then explain the method used to detect the produced atoms and reject the background, and how  the resulting signal is obtained.

\section{Experimental setup}
\begin{figure}[h]
\begin{center}
\includegraphics[width=10cm]{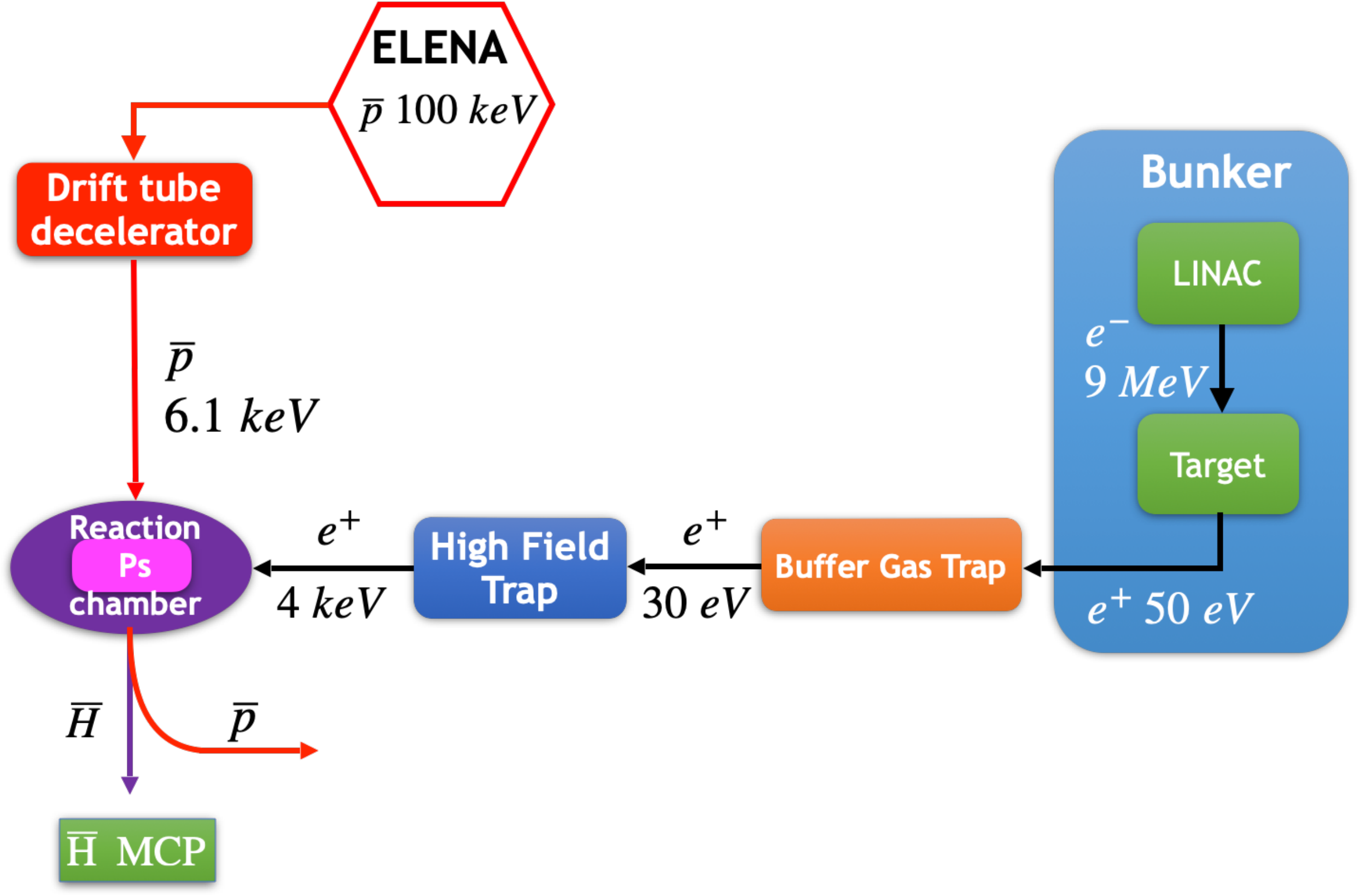}
\caption[]{\it{Scheme of the experiment. 
The positrons from the linac based source are  accumulated in a set of two traps and directed to the reaction chamber where they are converted into positronium. Antiprotons from ELENA  are decelerated with a drift tube and guided to the Ps target. The produced antihydrogen atoms are detected with an MCP while the antiprotons that did not react with Ps are deflected.}}
\label{fig:synoptic}
\end{center}
\end{figure}
A brief overview of the different parts of the apparatus is given here and illustrated schematically in Figure~\ref{fig:synoptic}. 
The antiprotons are delivered by the CERN AD-ELENA facility~\cite{ELENA} at an energy of 100~keV, typically every 110~s. 
The \pbar\ bunch is decelerated with a drift tube~\cite{DTpaper} and directed towards the Ps target located in the reaction chamber (RC). A linear accelerator produces electrons that impinge upon  a tungsten target equipped with a tungsten mesh moderator. 
The outcoming low energy positrons~\cite{pospaper} are guided to the Penning-Malmberg traps. 
A buffer gas trap (BGT) catches and cools the positron 
bunches. The \eplus\ are then transferred to a high field trap (HFT) where they are  accumulated~\cite{postrappaper}
between two \pbar\ pulses and ejected towards the reaction chamber.

In the RC the \eplus\ are converted into a cloud of ortho-positronium (oPs), the triplet spin state with a lifetime of 142~ns, which serves as the target for the \pbar\ to produce \Hbar\ atoms according to
reaction (\ref{eq1}).
The neutral anti-atoms hit a Micro Channel Plate (MCP) detector located in a straight line with respect to the incident \pbar\ beam. 
This is our main device to detect \Hbar. It will be denoted as \Hbar\ MCP in the following and described in detail in section~\ref{sec:hbarprod}. The antiprotons that did not participate to the reaction are deflected away from the detector.
Details on the production of the positronium target and on the antiproton beam are given in the two following sections.

\section{Production of the positronium target}

The positron beam is produced using a 9 MeV electron linear accelerator equipped with a water cooled tungsten target~\cite{pospaper}. The  peak electron current is 330 mA, with a pulse length of 2.85 $\mathrm{\mu}$s. The repetition rate can be varied from 2 to 300 Hz.  The high energy positrons (average 1 MeV) produced by pair creation in the 1 mm thick tungsten target are moderated to typically 3 eV (the work function of \eplus\ in W) in a series of tungsten meshes located beneath the target and accelerated to 50~eV for further transport in the beam line~\cite{pospaper}. The slow positron flux is  $\mathrm{2.9\times 10^{7}}$  per second.

\begin{figure}[h]
\begin{center}
\includegraphics[width=13cm]{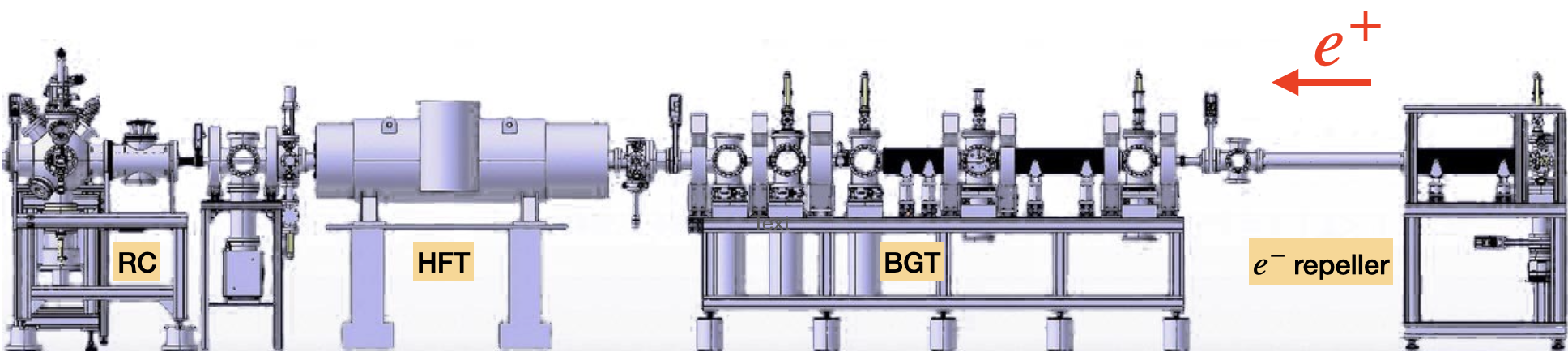}
\caption[]{\it{Side view of the 9.5~m long transport line of positrons from the linac (right, not shown) to the reaction chamber (RC at extreme left), with the electron repeller, the buffer gas trap (BGT) and the high field trap (HFT) in between.
}
}
\label{fig:BGT_RC_Transport}
\end{center}
\end{figure}

A schematic of the positron beamline with the key components is shown in Figure~\ref{fig:BGT_RC_Transport}.
The first element is an electrostatic repeller to reject electrons.
The BGT captures and cools the positrons using 
$\mathrm{N_2 \ (10^{-4} \; mbar\;pressure)}$ and $\mathrm{CO_2 \ (5 \times 10^{-5} \; mbar\;pressure)}$ in a magnetic field of the order of 40~mT. 
It contains three stages of coaxial cylindrical electrodes. We apply voltages in the 30-140~V range to form potential wells to trap the positrons and dynamically transfer them along this setup.
Positrons are transferred at 1~Hz repetition rate to the HFT where they are accumulated in a 5~T field in a vacuum better  than $10^{-10}$~mbar. 
The cooling time by synchrotron radiation in this field is 0.16~s. The electrode
stack consists of 27 electrodes in a 1.88~m long assembly. Typically 90 positron bunches from the BGT are accumulated by opening and closing the entrance potential barrier
of the potential well at the 1~Hz frequency. Details of the trapping schemes can be found in reference~\cite{postrappaper}.

The positron pulses, ejected at 300 or 500~eV from the HFT, are imaged with an MCP that can be moved in and out the beam line. 
The deposited charge can be measured using a fast charge sensitive pre-amplifier~\cite{CSPA}, while the front face of this MCP is biased at + 120 V to collect secondary electrons. 
The average number of positrons observed from the HFT was $1.5 \times 10^8$ per ELENA pulse.
The overall efficiency of
the system of the two traps is 5\%.

Ideally, the magnetic field in the antiproton transport line should be zero. However, the 5~Tesla magnet of the HFT produces a field of about 2~mT at the crossing of the beam lines in the RC. 
Thus a magnetic shielding box was made consisting of 3~mm thick soft iron around the target location, reducing the field intensity at the target from 2~mT to about 0.4~mT. An extra layer of iron bars was added to increase the thickness of the magnet return yoke of the HFT. This reduces the fringe field to about 0.2~mT in the interaction zone. 
The positron beam traverses the 3~mm shielding wall via a 40~mm diameter hole located at 161 cm from the trap centre. The transition to the field-free region is made non-adiabatically to prevent the divergence of the positron beam~\cite{Cooke}.
 
\begin{figure}[h]
\begin{center}

\includegraphics[width=13cm]{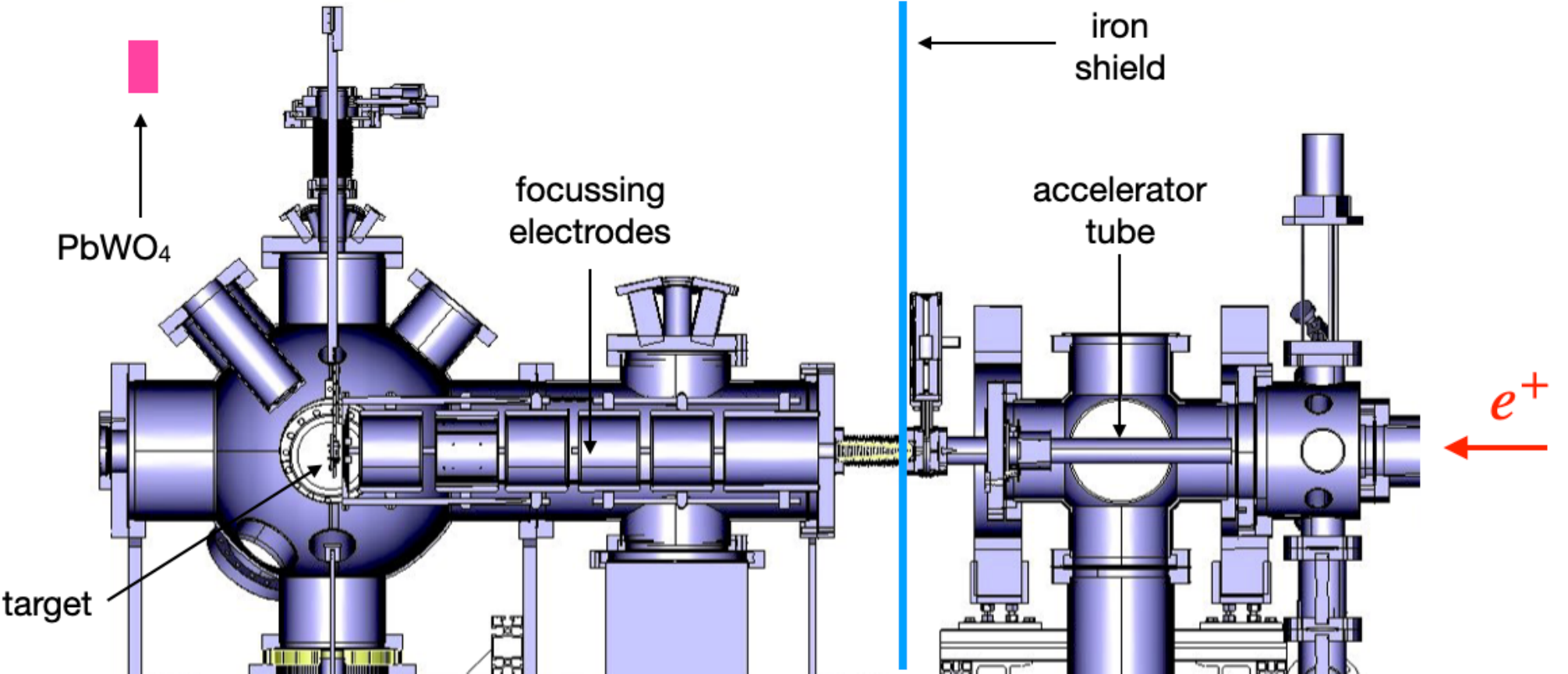}
\caption[]{\it{Side view of the transport of positrons to the Ps target. The HFT at the right ejects its stored positrons to the accelerator tube. The vertical blue line represents the iron wall of the magnetic shielding box. Six cylindrical electrodes focus the beam to the target position at the centre of the spherical reaction chamber where they are converted into Ps. A $PbWO_4$ crystal (pink) is placed close to the interaction point. 
}}
\label{fig:HFT_RC_Transport}
\end{center}
\end{figure}

 Just before this transition, the positrons are further accelerated with a switched drift tube accelerator so that their final energy is 4.3~keV. This energy facilitates the transport into the magnetic field free zone where the positron beam  is focused using a set of 100~mm diameter cylindrical electrodes (Figure~\ref{fig:HFT_RC_Transport}) and two planar electrodes to shape the beam so as to optimize the overlap with the antiproton beam profile in the interaction region. At the beam axes crossing point in the reaction chamber, the positron bunch is 17 ns (FWHM) long and its arrival time is set 30~ns earlier than the arrival time of the \pbar\ bunch. This delay between particle bunches was found in simulations to be optimal for  \Hbar\ production.

A sample holder, located at the centre of the RC, holds a flat
$\mathrm{19~mm \times 19~mm}$ plate made of conductive silicon single crystal, on which a nanoporous silica film was deposited~\cite{Psconverter,Psconverter2}. At 4.3~keV the positrons are implanted deep enough in this Ps converter for ortho-positronium (oPs) to be emitted at low energy. Only oPs with a 142~ns lifetime has a chance to interact with an antiproton for \Hbar\ production whereas para-positronium, which has a lifetime of 125~ps, does not leave this film.

\begin{figure}[h]
\begin{center}
\includegraphics[width=13cm]{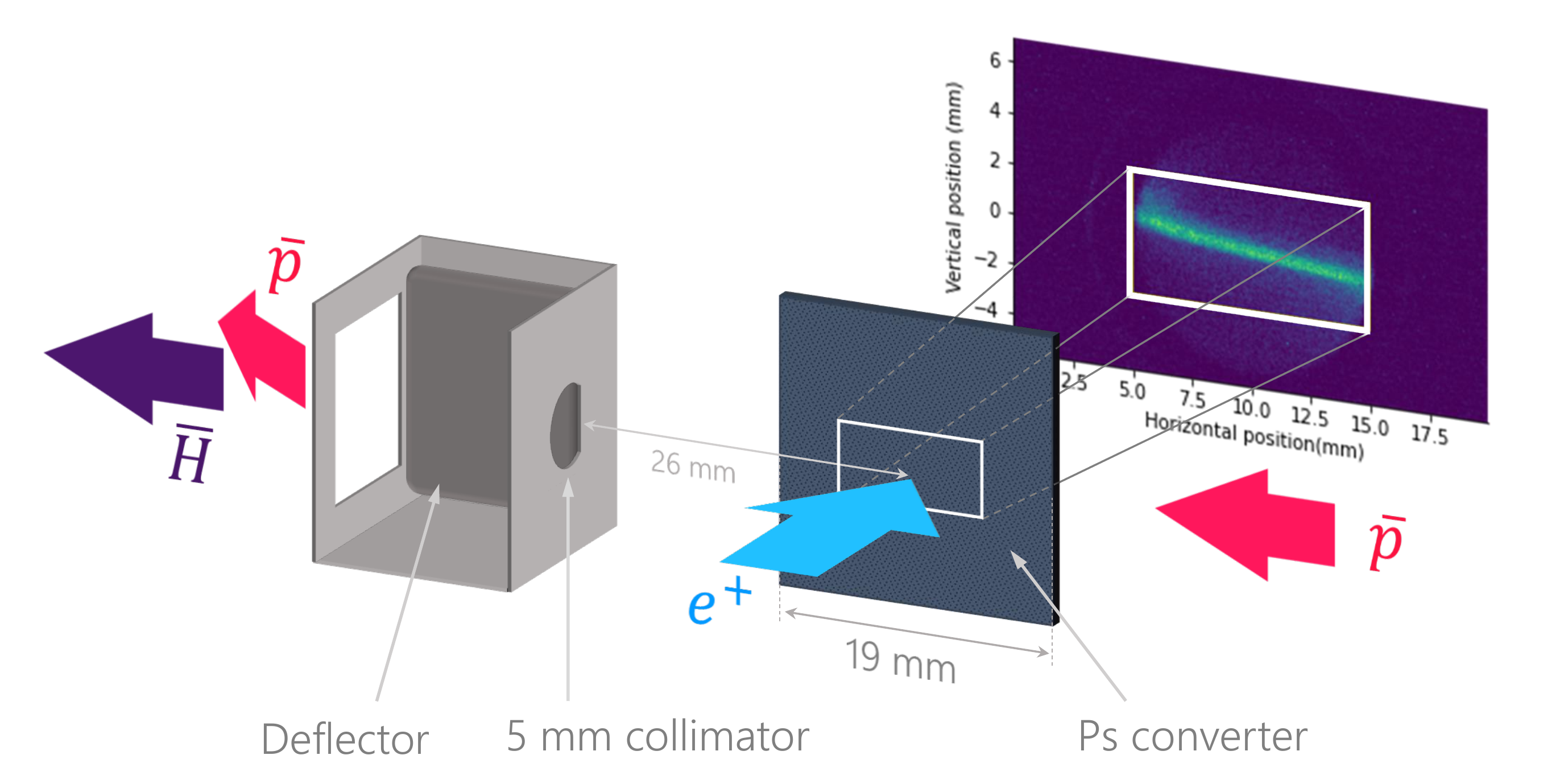}
\end{center}
\caption[]{\it{
Scheme of the target region. An electrostatic deflector is located on the beam axis with its 5~mm diameter aperture 26~mm downstream the target centre acting also as a collimator. The positron beam profile seen by the MCP when placed at the same location as the Ps converter} is also  shown.The white rectangle figured on the screen has a length of 10 mm and a height of 5 mm.}
\label{fig:pos-spot-RC}
\end{figure}

On the sample holder, another plate of the same dimensions as the Ps converter can be placed at the same location for reference.  It is made of silicon and does not produce oPs. 
At a third position on the sample holder, an image sensitive 10~mm diameter MCP can also be moved to  intersect the positron beam to check its focusing as sketched in Figure~\ref{fig:pos-spot-RC}.
Using this MCP, the number of positrons reaching the RC is measured in a similar way as at the exit of the HFT with the same charge sensitive pre-amplifier. On average $5.2 \pm 0.5 \times 10^7~e^+$ per pulse reaches the target plane,
i.e. a 35\% transport efficiency,  with variations depending on the running conditions from 4 to 7~$\times 10^7~e^+$. 
A measure of the overlap of the positron bunch with the antiproton beam is given by the fraction of positrons in a rectangle
of 19~mm (the length of the Ps converter) by 5~mm (the diameter of the aperture of the collimator  shown in Figure~\ref{fig:pos-spot-RC}), which is  $73 \pm 3$~\%.

\begin{figure}[h]
\begin{center}
\includegraphics[width=8cm]{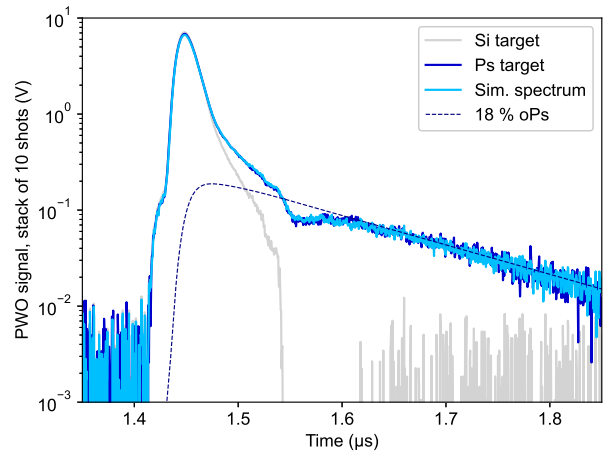}
\caption[]{\it{SSPALS spectrum obtained with a $PbWO_4$ crystal. The dip at $\mathrm{1.55 \: \mu s}$ is due to the detector response. The data from the Si (grey) target is used to generate a template SSPALS (light blue) including ortho positronium formation and decay. An oPs fraction of 18 \% (dashed) provides a good fit for the experimental data of the mesoporous silica target (dark blue).}
}
\label{fig:Ps_lifetime}
\end{center}
\end{figure}

A $\mathrm{PbWO_4}$ crystal detects the gamma rays emitted by the annihilation of the positrons hitting the target plane and by the decay of oPs. To estimate the number of oPs produced, we use the SSPALS technique~\cite{SSPALS}, a method to measure positronium lifetimes from a single intense burst of positrons. We find that $18 \pm 3$~\% of those positrons are converted into oPs (Figure~\ref{fig:Ps_lifetime}), resulting, on average, in $6.8 \pm 1.5 \times 10^6$ oPs created in 
the $\mathrm{19 \: mm \times 5 \: mm}$ rectangle.

This is thus the number of Ps atoms created for each ELENA pulse within a volume delimited by the collimator and this rectangle. In the next section we describe how we prepare the \pbar\ component for the reaction.

\section{Antiproton beam}
\label{sec:pbar}
We use a drift tube to decelerate the antiprotons 
from 100 keV, as received from ELENA, to  energies below 10 keV, suitable for \pbar\ trapping and producing antihydrogen~\cite{DTpaper}. 
This tube is held at more than 90~kV for 3~s before the antiproton bunch arrives, and is then switched to ground in  18~ns while the particles are inside. A leakage current varying between 10 and 30 $\mu$A on a 5 M$\Omega$ resistor leads to a voltage drop between 50 and 150~V that modifies the final beam energy. 
With a high voltage setting at 94 kV, the decelerated beam has an average energy of $6.10 \pm 0.05$ keV. 

During the 2022 run, the intensity at extraction from ELENA was progressively increased from about $5 \times 10^6$ to about $7 \times 10^6$ \pbar\ per bunch. The absolute accuracy of those values is of the order of 20\%.
Thanks to bunch rotation~\cite{bunchrotation,bunchrotation2}
performed in ELENA, the 100~keV antiproton bunch has a length of 40~ns (RMS) and
a time-jitter of 4~ns. With these values, the bunch is well contained in the 450~mm long deceleration drift tube. 
This means that practically 100\% of the \pbar\ bunch is decelerated. The bunch rotation implies a doubling of the energy spread, up to $2 \times 10^{-3}$, but this has a negligible impact. The horizontal and vertical emittances were carefully measured by the ELENA team as 2.9 and 2.1 $\mathrm{mm \times mrad}$ (RMS) respectively. 
These values are about 2.5 times larger than design, with a significant effect on the decelerated beam size and on the possibility to focus it at the point of reaction with the Ps.
The bunch length after deceleration is 100~ns (RMS).
ELENA also enabled the delivery of 100 keV \Hminus\ ions every 15~s that were used to perform a first adjustment of the electrostatic optical elements of the beam line. 
Plastic scintillator detectors are placed along the beam line. Their time resolution of a few ns allows to locate the places where antiprotons annihilate and helps adjusting the voltages of the optical elements (steerers and Einzel lenses) to optimise \pbar\ transmission. 

\begin{figure}[ht]
\begin{center}
\includegraphics[width=13cm]{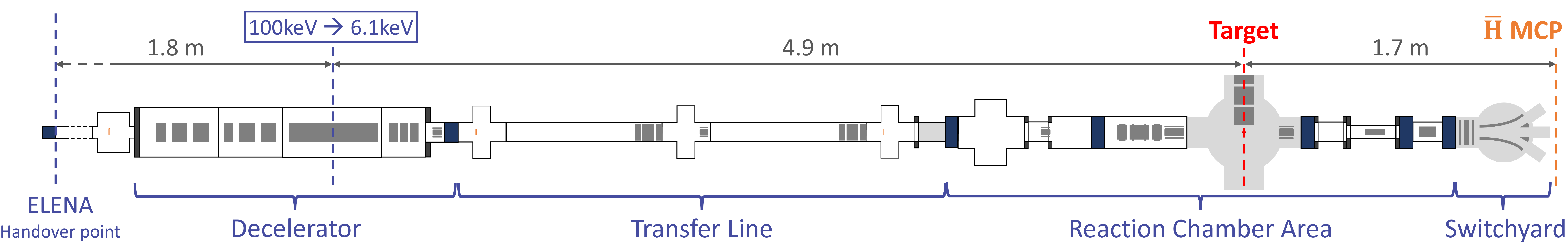}
\caption[]{
\it{
GBAR $\overline{p}$ beam line in 2022. The beam is decelerated by the drift tube, then focussed in the reaction chamber where it meets the Ps target to form $\overline{H}$ atoms. After the RC an electrostatic switchyard can deflect the $\overline{p}$ beam to the ion beamline, where a $\overline{p}$ trap (not shown) is currently located. For the $\overline{H}$ formation experiment, the deflector located right after the reaction region removes most of the $\overline{p}$s, while the switchyard removes the remaining ones. The neutral atoms are detected in the $\overline{H}$ MCP located in straight line downstream.}}
\label{fig:pbarline}
\end{center}
\end{figure}

A SIMION{\textsuperscript{\textregistered}} 8.1~\cite{SIMION} simulation, interfaced with COMSOL{\textsuperscript{\textregistered}}~\cite{COMSOL} field maps,  
models the transport and focussing elements of the \pbar\ beam to the reaction chamber. 
This simulation, using the 100~keV ELENA beam parameters, allows to estimate that the fraction of the \pbar\ bunch that intersects the 5~mm collimator located after the Ps converter in the RC is 38\%. 
The divergence of the neutral beam results in a fraction of the particles not hitting the \Hbar\ MCP, with an estimated loss of 32\%.

The average number of 6~keV antiprotons per bunch is estimated in special runs where the deflectors after the target are not activated, and the antiprotons are stopped in the \Hbar\ MCP. 
A CMOS sensor was employed in an externally mounted commercial digital camera~\cite{BAUMER} without lens located close to the MCP.
Thanks to the large number of pixels (5M) and a detection efficiency close to 100\%, the sensor can reliably count charged particle tracks in the high density environment generated by the annihilation of millions of antiprotons. The number of annihilations is inferred from the number of tracks, taking into account the number and nature of annihilation products per antiproton, which depends on the nature of the target atoms \cite{anucnucia}, and the effect of material between the target and the sensor, derived from a simple GEANT model~\cite{Geant4}).
The uncertainty of the method is estimated to be around $\pm 25\%$ mainly due to different materials on the MCP surfaces and a rough assessment of the effect of the material of the beam pipe. 
The number of antiprotons reaching the \Hbar\ MCP, with the \pbar\ deflection electric fields off (see section~\ref{sec:hbarprod}), is determined to be about $2.3 \times 10^6$ per spill. Taking into account a  transmission efficiency of  74\%, from simulation, we estimate the  number of antiprotons passing through the positronium cloud to be about $3.1 \times 10^6$ per spill on average. This number and the number of positroniums obtained in the previous section  serve as inputs to estimate the expected number of antihydrogen atoms produced according to reaction~(\ref{eq1}) as described in the following section.

\section{\Hbar\ production measurement}
\label{sec:hbarprod}
\begin{figure}[h]
\begin{center}
\includegraphics[width=13cm]{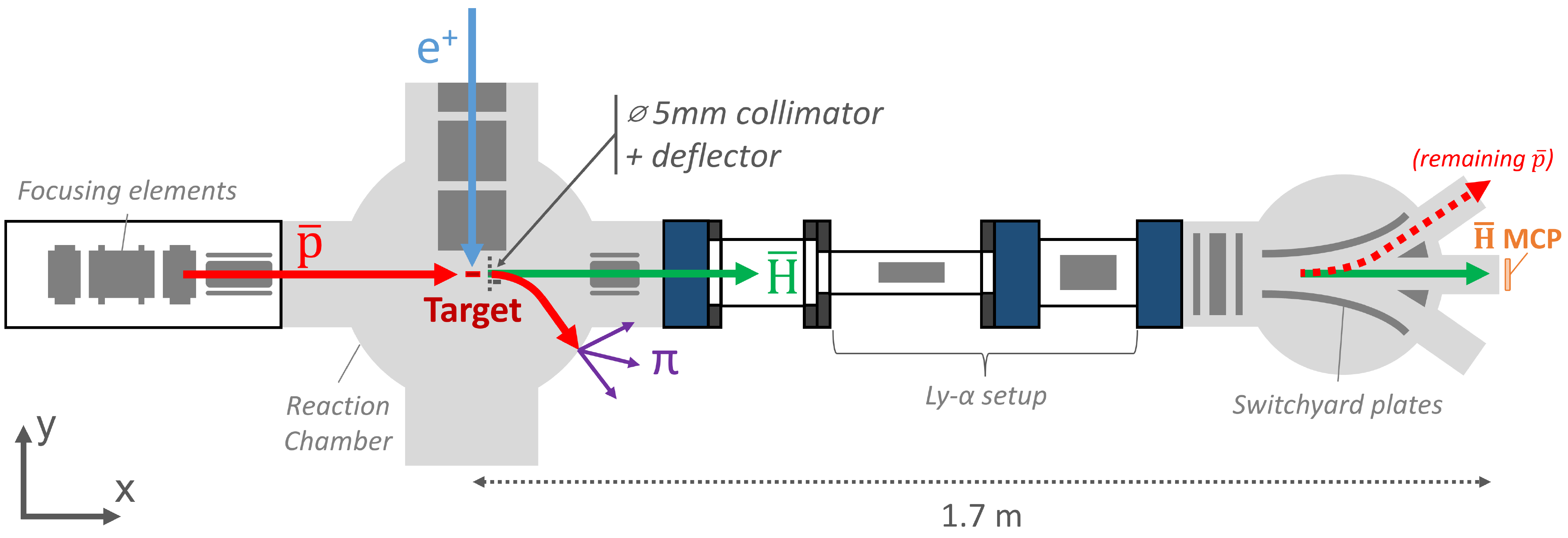}
\caption[]{\it{Sketch of interaction and detection scheme.}}
\label{fig:RC_area}
\end{center}
\end{figure}

The parts of the experiment where the creation of antihydrogen and its detection take place are shown schematically in Figure~\ref{fig:RC_area}.
The antiprotons are focussed at the target location where they meet the positronium cloud. The antiprotons that did not react and emerge from the Ps cloud are directed towards the walls of the RC by applying a voltage on an electrostatic deflector whose 5 mm diameter aperture acts as collimator (see Figure~\ref{fig:pos-spot-RC}). The neutrals fly undeflected towards the  \Hbar\ MCP detector located on the \pbar\ beam axis 1.7 m downstream the Ps target. In case that some antiprotons, for instance from the beam halo, survive the deflector, the electric field in the switchyard electrodes deviates them towards a beam line at 35 degrees from the main axis.

 The \Hbar\ MCP is composed of a set of two Micro Channel Plates in chevron configuration from Photonis (Advanced Performance Detector model), with a gain of $1.76 \times 10^7$, at 2200~V, followed by a fast P46 phosphor screen (about 100~ns decay time). The MCP has a diameter of 40~mm and a thickness of 1~mm. The time resolution for a single particle is less than 2~ns. A  camera, model pixelfly from PCO with Sony ICX285AL image sensor, with a resolution of 1392 x 1040 pixels, records the image from the phosphor screen with a shutter speed set at $\mathrm{1 \: \mu s}$.

To estimate the expected production of antihydrogen, the time evolution of  Ps diffusing out of the silica target is taken into account as well as the time and spatial overlap of the interacting particles. 
Ps is assumed to be emitted from the target with a distribution corresponding to the implanted positron profile measured by the position sensitive MCP at the target position (see Figure~\ref{fig:pos-spot-RC}), which is well represented by a two dimensional  profile of widths 3.5 and 0.5 mm along the antiproton beam axis and the vertical axis respectively.
The positron pulse length is 17~ns (FWHM).
The implantation energy  of 4.3 keV in the  nanoporous thin film results in a delay in the emission of oPs of 10~ns with 2~ns RMS~\cite{Psdiffusion}. 
The velocities of the Ps atoms can be described with a Maxwell-Boltzmann distribution at 600~K emitted with a cosine distribution ($\mathrm{dn/dcos(\theta) \propto cos(\theta)}$)~\cite{Psvelocity1}, as validated by experiments at ETH Zurich~\cite{Psvelocity}. 
The 5~mm diameter collimator located on the \pbar\ beam axis 26~mm after the positron converter defines the transverse acceptance for the anti-atoms. The target length along the antiproton axis and the collimator transverse to it, define the volume in which the Ps and \pbar\ interact. 
The calculated cross section is dominated by  2P states which after production would decay promptly (1.6 ns) to the ground state. 
In the Close Coupling method~\cite{Kadyrov} the predicted value is $\mathrm{13.4 \times 10^{-16}~cm^2}$ at 6.1~keV incident \pbar\ energy, while it is  $\mathrm{30.6 \times 10^{-16}~cm^2}$ in the Continuum Distorted Wave approach~\cite{Comini}. The expected production rates are $1.3 \pm 0.4$ \Hbar\  and $3.0 \pm 0.9$ \Hbar\ per 100 spills for the two models respectively, the uncertainty coming from the \pbar\ flux and Ps number. 

The data were taken from October to November 2022, in periods of 8~h shifts. The ``MIX" periods combined the antiprotons with positronium (6897 spills), while the ``BGD" data were taken without positrons (8468 spills). Data were also recorded with only positrons and no antiproton beam to study the background due to positron annihilation or Ps decay. Some data were also collected with no beam to study the possible backgrounds due to the environment and cosmic rays. 

\begin{figure}[h]
    \centering
    \includegraphics[width=0.6\linewidth]
    {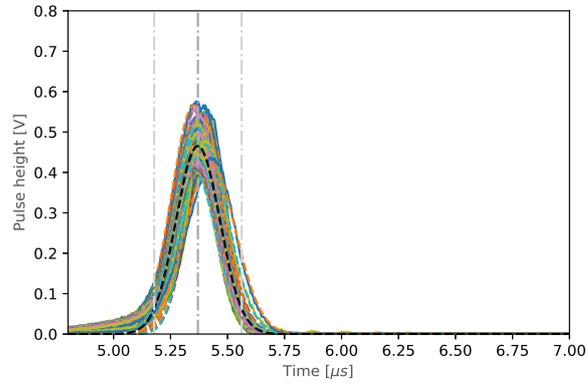}
    \caption{Electric signals of multiple \pbar\ bunches on \Hbar\ MCP. The spread is generated by the longitudinal bunch profile and well described by a Gaussian distribution with a 96~ns FWHM. The vertical lines define the signal time window: $\mathrm{5.37  \pm 2 \times 0.096 \; \mu s}$.    }
    \label{fig:pbar_tof}
\end{figure}
\subsection{MCP electrical signal analysis}
At 6.1 keV, the time of flight between the creation point in the reaction chamber and the detector is 1.6~$\mu$s. 
The bunch of antiprotons produces a signal of 96 ns width (RMS) on the \Hbar\ MCP when the electrostatic deflector is grounded, as shown in Figure~\ref{fig:pbar_tof}.  
A time interval of $\pm 2$ standard deviations around the central arrival time of these undeflected antiprotons (Figure~\ref{fig:pbar_tof}) is defined as the time span during which we accept \Hbar\ candidate events. 

\begin{figure}[ht]
    \centering
    \begin{minipage}{0.49\textwidth}
	\includegraphics[width=6.5cm]
            {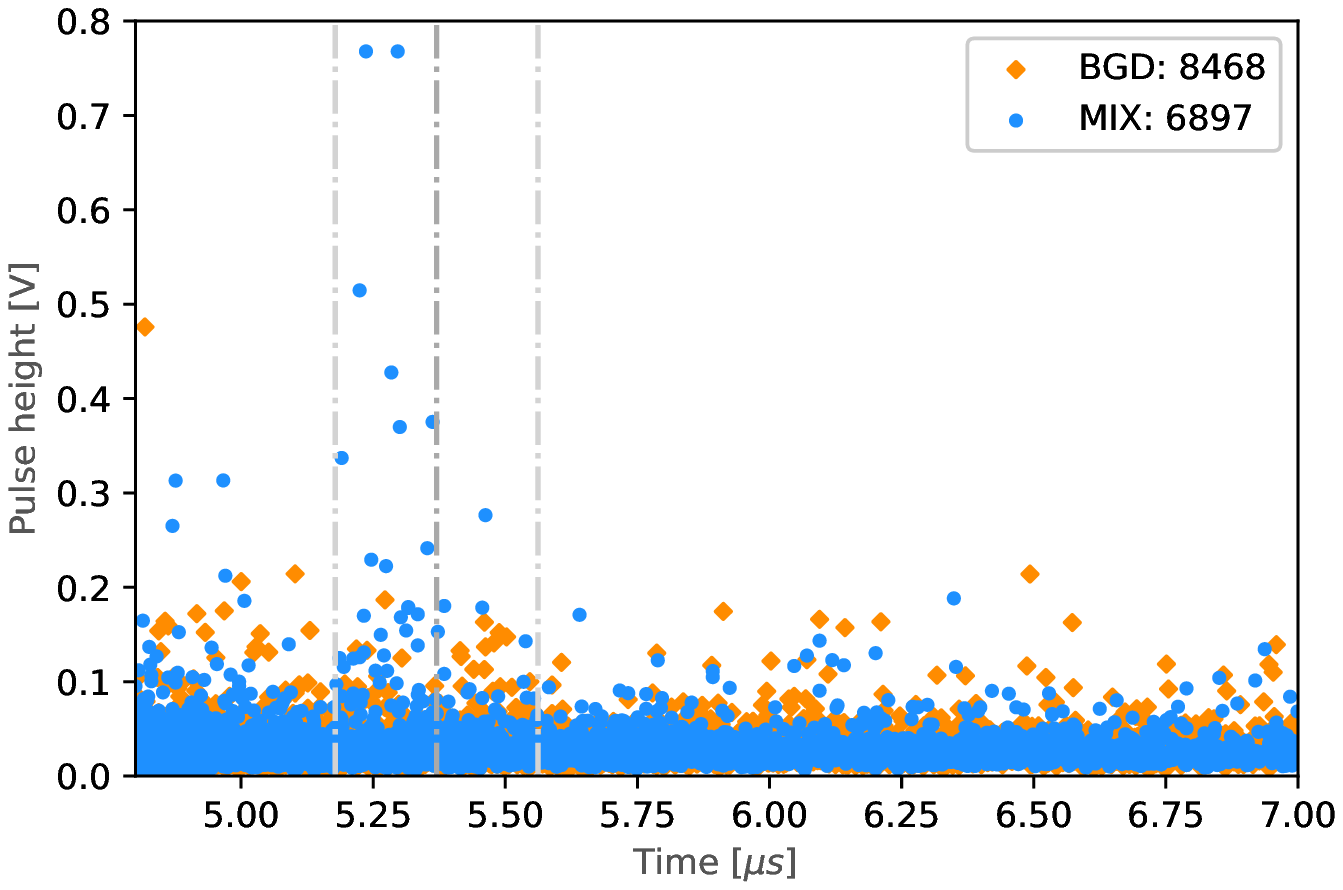}
    \end{minipage}
    \hfill
    \begin{minipage}{0.49\textwidth}
	\includegraphics[width=6.5cm]
         {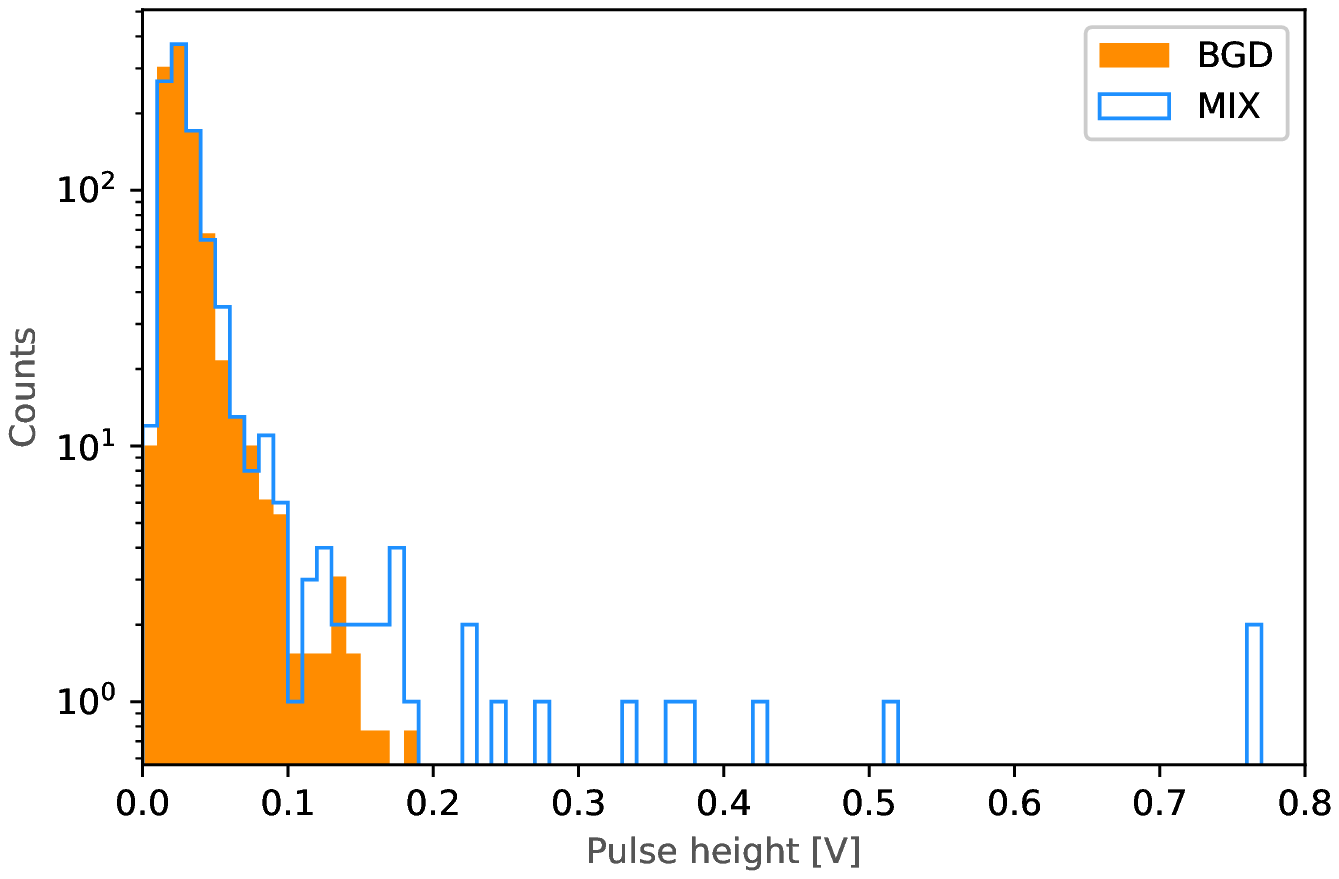}
    \end{minipage}
\caption[]{\it{Left: Pulse-height ($V_{max}$) vs. time ($T_{max}$) distribution of MCP signals for MIX events (blue) and for BGD events (orange). The vertical gray lines indicate the time-of-flight signal region  where $\overline{H}$ atoms are expected to annihilate on the detector. 
Right: Pulse height ($V_{max}$) distribution of MIX events (blue) and BGD events (orange) in the signal time-interval. The BGD histogram is scaled to the number of MIX spills.}}
\label{fig:elec-dist}
\end{figure}

The time width of the recorded signals is typically smaller than 2 ns. Several such signals are recorded by the MCP per shot in the 10 $\mu$s time window of the data acquisition system. We define  $\mathrm{V_{max}}$  as the maximum voltage of an event and $\mathrm{T_{max}}$ its associated time. Figure~\ref{fig:elec-dist} shows the 2D distribution of $\mathrm{V_{max}}$ versus $\mathrm{T_{max}}$ for the MIX and BGD samples.  In the MIX data there are clearly events with a large pulse height in the signal time-window, which are not present in the BGD data. This is what is expected for antihydrogen formation since in this case anti-atoms can reach and annihilate on the MCP while antiprotons are deflected and annihilate upstream, thus only pions can reach the MCP.

The gamma rays from oPs decay may also be a source of background. This is largely reduced with the selection of events in the signal window, corresponding to 10 lifetimes of oPs. Given the MCP solid angle and a detection efficiency of 5\% that takes into account the high gain ($10^7$), a total of 18 such photons would have been detected.  
Considering the MCP efficiency, measured with photons from a $\mathrm{^{22}Na}$ source, of less than 1\%, after a cut on $\mathrm{V_{max}}$ at 0.05~V, this background is negligible.
   
Figure~\ref{fig:elec-dist} shows the distribution of $\mathrm{V_{max}}$ for candidate events in the MIX and BGD runs in the signal window. 
Testing for the presence of a signal amounts to testing that the Poisson distributions which generated the MIX and BGD samples have the same average parameter. 
Let $\mathrm{N_{mix}}$ and $\mathrm{N_{bkd}}$ be the total number of events and $\mathrm{n_{mix}}$ and $\mathrm{n_{bkd}}$  the number of observed signal-like and background-like events. If they come from the same law, $\mathrm{n_{mix}}$ and $\mathrm{n_{bkd}}$ should be shared along a binomial distribution whose parameter p is approximated by :
$\mathrm{p = \frac{Nmix}{N_{mix}+N_{bkd}}}$~\cite{stats}. The binomial test calculates the (two-sided) probability that the observed 
$\mathrm{n_{mix}}$ departs from its natural value 
$\mathrm{p \times (n_{mix} + n_{bkd})}$. This probability is then translated into a number of standard deviations by the usual formula for Gaussian distributions.  The  significance varies with the cut value between 3.1 and 4.1 standard deviations for cut values above 0.05 V. 

For a minimum voltage cut of 0.1 V, the number of events in MIX  and BGD shots are 32 and 15 respectively. The normalised background is $12.2 \pm 3$, i.e. an excess of $19.8 \pm 6$ with a significance of  3.1 standard deviations. The efficiency of the detection by the MCP for this cut value is estimated to be about 50\%. 
The nominal acceptance of \Hbar\ atoms in the detection area of the MCP is estimated by simulation to be 68\% (see section~\ref{sec:pbar}). 
However we observed perturbations of the charged beam with respect to its nominal trajectory, which might impair the acceptance for the neutrals, hence this value can only be taken as an upper limit. 
Taking into account the number of shots in the MIX and BGD samples, and the efficiencies and acceptances described above, the excess in the MIX shots corresponds to $8.9 \times 10^{-3}$ \Hbar\ produced in the Ps target per antiproton pulse.
Given the large uncertainty in the acceptance of \Hbar\ atoms, this number is in a rough agreement with the number between $1.4 \pm 0.4 \times 10^{-2}$  and $3.2 \pm 0.9 \times 10^{-2}$ \Hbar\ per antiproton pulse, expected from different theoretical models. In the future, several improvements will be brought to the control of the charged and neutral trajectories to enable a proper cross-section measurement. 

\subsection{MCP image  analysis}

\begin{figure}[h]
\centering
\begin{minipage}{.5\textwidth}
  \centering
  \includegraphics[width=1\linewidth]{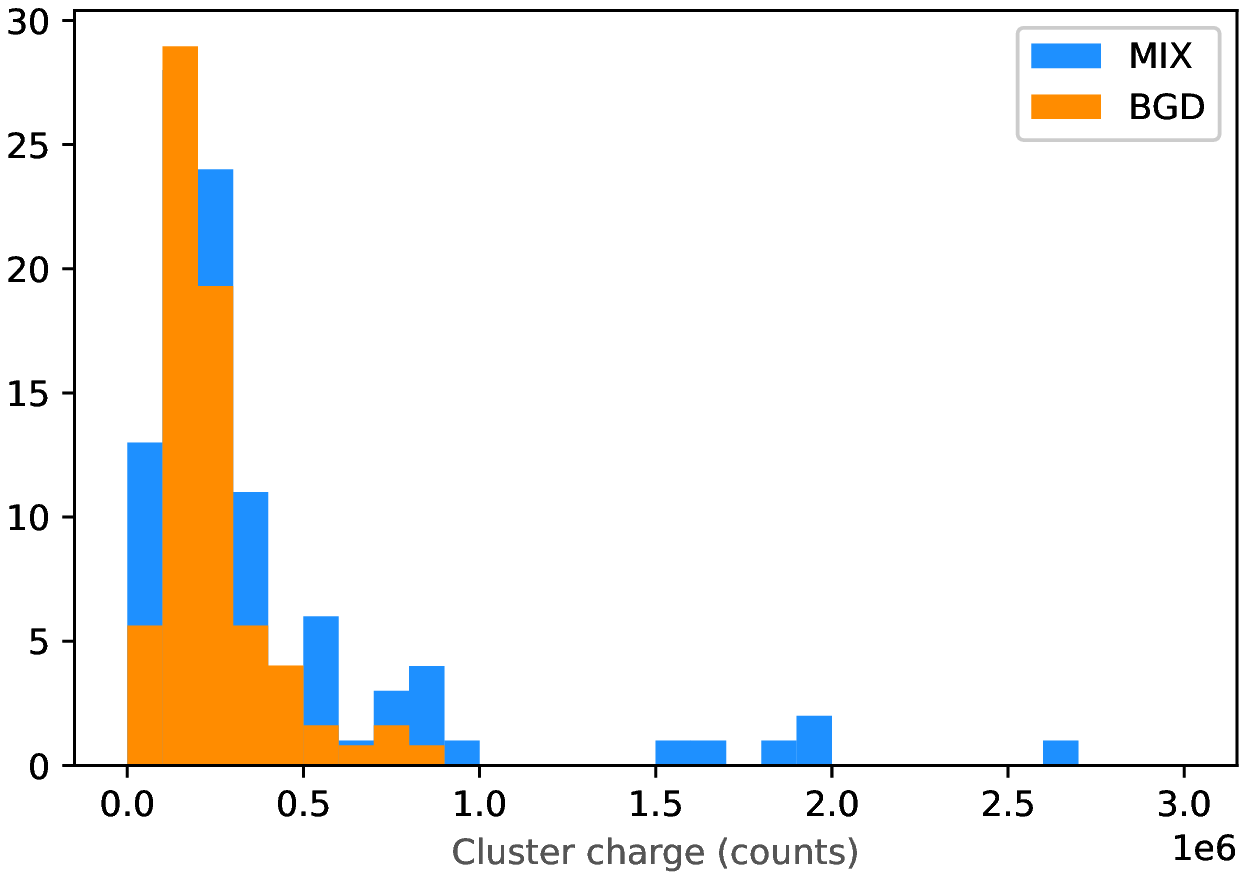}
\end{minipage}%
\begin{minipage}{.5\textwidth}
  \centering
  \includegraphics[width=1\linewidth]{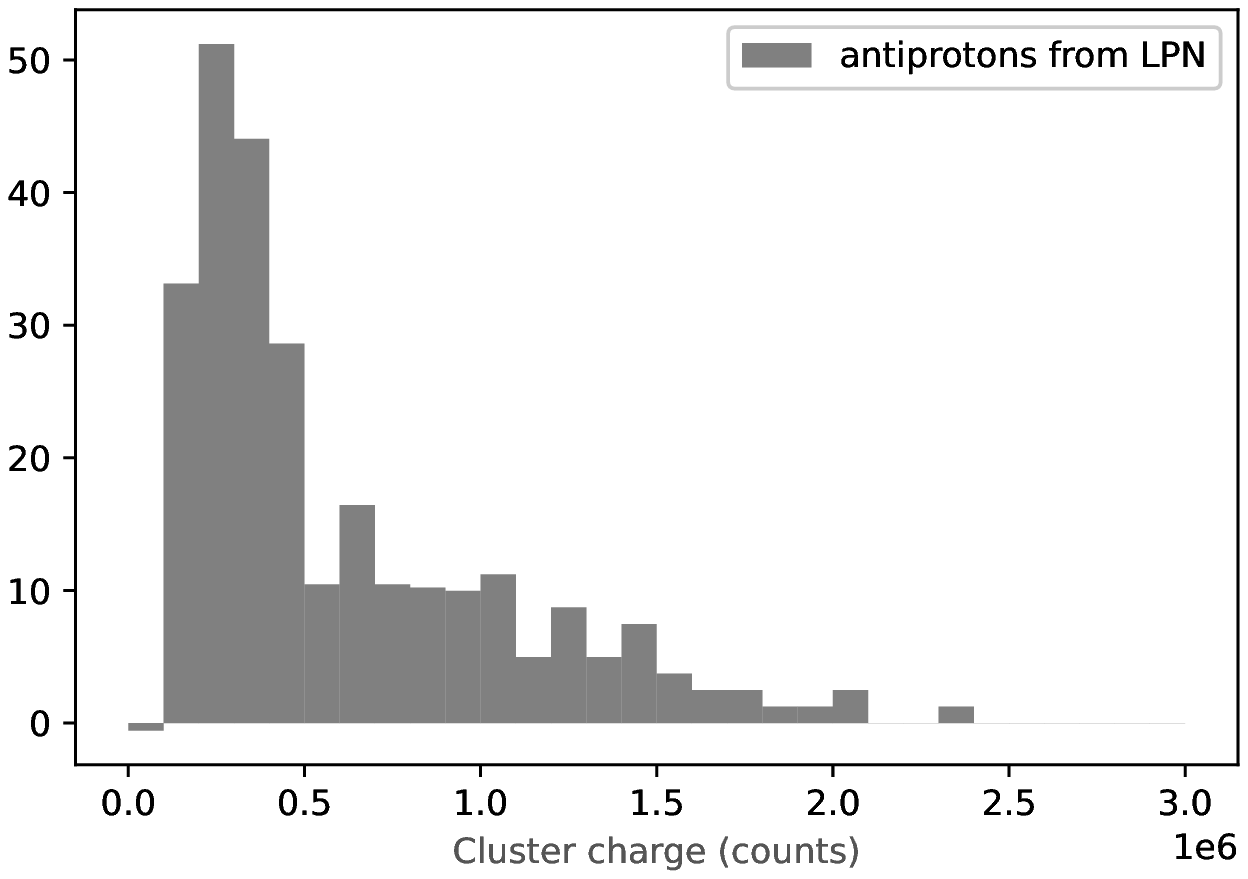}
\end{minipage}
  \caption{Distribution of Cluster charge. Left: MIX and BGD shots, right: antiprotons (from LPN shots)}
\label{fig:clusters}
\end{figure}

\begin{figure}[h]
    \centering
    \includegraphics[width=8cm]
    {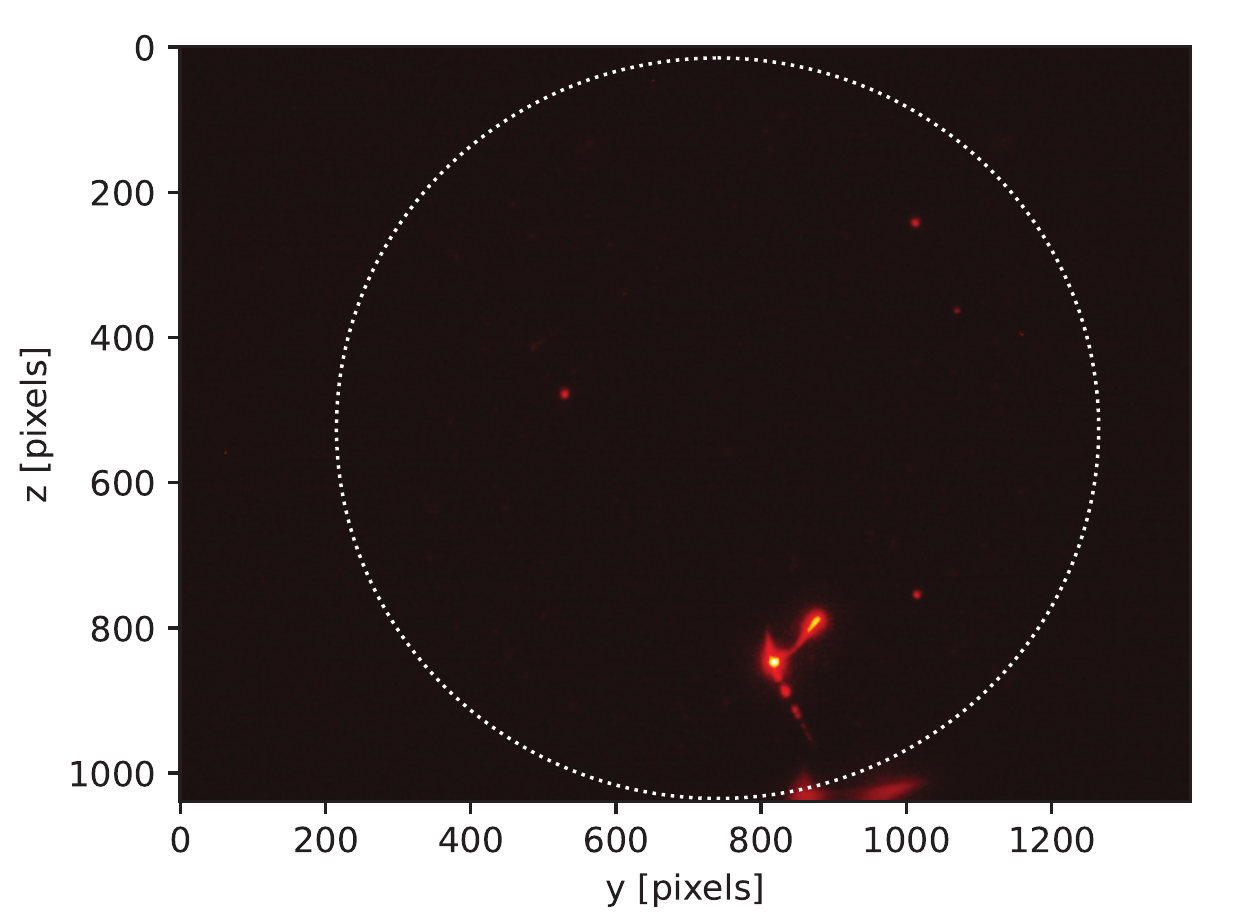}
    \caption{Example of  MCP image for a large charge MIX event. The dotted circle represents the limit of the MCP active area.}
    \label{fig:MCP_image}
\end{figure}

A cross-check of the analysis can be performed using the MCP images recorded for the same spills. The MCP image for the impact of an \Hbar\ atom is expected to be similar to that for the impact of an antiproton. The images display individual impacts, however it is not possible to attribute both a precise timing and a precise charge to a single impact, since the screen and the camera integrate over a time interval of $1\; \mu s$. This is in contrast with the electrical pulse from the MCP, used in the analysis above, which is very short but  integrates over the whole MCP area. To characterize the image of these impacts, special runs (LPN, for Low \pbar\ Number) were taken with a low number of antiprotons reaching the MCP, by detuning the incoming \pbar\ beam. In these shots, the average hit multiplicity on the MCP is 7.5 with large fluctuations and containing a substantial number of pions produced by the annihilations of the antiprotons deflected upstream.
The images are analysed with the help of a simple clustering algorithm. The charge of a cluster is defined as the sum of the collected charge from each  pixel belonging to it.
The distribution of the cluster charge for antiprotons is extracted from the difference between the  distributions for LPN and Background runs 
and is shown in Fig~\ref{fig:clusters} (right).
 Images in the MIX and BGD samples are analysed with the same clustering algorithm. The resulting distributions of cluster charge are shown in Fig~\ref{fig:clusters} (left).
 The large charge events seen in the MIX sample have a cluster charge compatible to that expected from antiproton impacts, as extracted from the LPN data. A typical large-charge MIX event image is shown in Fig~\ref{fig:MCP_image}. Selecting clusters of charge superior to $5 \times 10^5$ counts  keeps 50\% of them. Using this minimum value, the number of events in MIX and BGD runs are 22 and 6 respectively. The normalised background is $4.8 \pm 2$ i.e. an excess of $17.2 \pm 5.1$ events, with a significance of 3.6 standard deviations, similar to the results obtained with the electrical signal analysis.
  
\section{Summary and outlook}
We have observed the production of antihydrogen atoms via charge exchange of 6.1~keV antiprotons with positronium atoms in the fundamental state.

Theory predicts most of these anti-atoms to be in the ground state while they travel through the Ps cloud, which is beneficial for the subsequent production of \Hbarplus\ ions.  In the future, GBAR intends to produce \Hbarplus, cool them  to velocities of the order of $\mathrm{\sim 1 \; m s^{-1}}$, and photo-ionize them to obtain slow \Hbar\ atoms.  The free fall of these atoms could then be measured by a classical time/position measurement, or even more accurately by a quantum interference method~\cite{quantuminterf}. The observed production of  6 keV \Hbar\ atoms constitutes the first milestone of this roadmap. Furthermore, in the presently demonstrated scheme of \Hbar\ production, a substantial fraction  of the order of 15\% of the anti-atoms should be in the 2S state, allowing to measure the Lamb shift of antihydrogen in  different conditions from those of the previous measurement~\cite{ALPHA-Lambshift}, in particular in a non-magnetic environment~\cite{Lambshift}. Such a measurement would be complementary with other efforts to search the CPT-violating parameter space for signs of new physics~\cite{Kostelecky}.

The rate of \Hbar\ production must be increased substantially in order to produce anti-ions. Several improvements are being considered on the positron line, such as reaching the full power of the linac, optimising the positron moderator geometry, implementing a new trapping scheme in the BGT by replacing the $\mathrm{N_2}$ gas by a solid state SiC re-moderator~\cite{Leite_2017}, and working on the transfer efficiencies between devices. On the antiproton line, a Penning trap will soon be added after the drift tube decelerator, with the possibility of electron-cooling the antiprotons. This should improve the emittance of the \pbar\ beam and allow to use a Ps confining cavity. 

\bmhead{Acknowledgments}
We thank F. Butin and the CERN EN team, L. Ponce and the AD/ELENA team for their fruitful collaboration. 
A. Prost and T. Stadlbauer from CERN and D.C. Faircloth from RAL are also warmly thanked for their help on HV techniques for the decelerator as well as A. Sinturel for help and expertise on vacuum. 
This work is supported by:
JSPS KAKENHI Grant-in-Aid for Scientific Research A 20H00150 and Fostering Joint International Research A 20KK0305 (Japan),
ANTION ANR-14-CE33-0008 (France), Programme National Gravitation, Références, Astronomie, Métrologie (PNGRAM), Institut de Physique, CNRS (France), the Swiss National Science Foundation (Switzerland) grants 197346 and 216673 and ETH Zurich (Switzerland) grant ETH-46 17-1, the Swedish Research Council (VR) grants 2017-03822 and 2021-04005 and the following grants from Korea: 
IBS-R016-Y1, 
IBS-R016-D1, 
UBSI Research Fund (No. 1.220116.01) of UNIST, 
NRF-2016R1A5A1013277, 
NRF- RS-2022-00143178,
NRF-2021R1A2C3010989, 
and 
NRF-2016R1A6A3A11932936. 
The GBAR collaboration is an International Research Network, supported by CNRS, France.

\bibliography{sn-bibliography}

\end{document}